\documentclass[times,twocolumn,final,nopreprintline]{elsarticle}

\usepackage{jasr}
\usepackage{framed,multirow}

\usepackage{amssymb}
\usepackage{latexsym}

\usepackage[switch]{lineno}

\usepackage{url}
\usepackage{xcolor}
\definecolor{newcolor}{rgb}{.8,.349,.1}

\usepackage[citebordercolor=white]{hyperref}

\usepackage{amsmath}

\hypersetup{
    colorlinks=true,
    linkcolor=cyan,
    filecolor=magenta,      
    urlcolor=cyan,
    pdftitle={Overleaf Example},
    pdfpagemode=FullScreen,
    }

\journal{Advances in Space Research}

\begin{document}

\verso{Given-name Surname \textit{etal}}

\begin{frontmatter}

\title{Numerical study of multiple solar flare induced modulation of Very Low Frequency (VLF) diurnal profile}

\author[1]{Sourav Palit\corref{cor1}}
\cortext[cor1]{Corresponding author: 
  Tel.: +91-9051639237;}
\ead{souravspace@gmail.com}
\author[1,2]{Subhajit Bhattacharyya}
\author[1,3]{Taraknath Bera}
\author[1,3]{Sandip K. Chakrabarti}

\affiliation[1]{organization={Indian Centre for Space Physics},
                addressline={466 Barakhola, Mahakash Sarani, Netai Nagar},
                city={Kolkata},
                postcode={700099},
                state={West Bengal},
                country={India}}

\affiliation[2]{organization={Dhulauri High School},
               addressline={Lalgola, Diar Fatepur},
               city={Murshidabad},
               postcode={742148},
               state={West Bengal},
               country={India}}

\affiliation[3]{organization={Ionospheric and Earthquake Research Centre \& Optical Observatory},
               addressline={Sitapur, Daspur},
               city={West Medinipur},
               postcode={721154},
               state={West Bengal},
               country={India}}


\begin{abstract}
\noindent Earth's ionosphere is a perpetual detector of ionizing radiation received from celestial objects, particularly the Sun. Solar ionizing radiation in the form of extreme ultraviolet (EUV) and X-rays during both quiet and active phase of the Sun, and charged particles associated with a solar wind imprint their ionization signatures on the ionosphere. The ionization rate varies on a timescale from a few minutes, for small flares, up to a few years, following the solar cycle, and also depends on geographical location and altitude. Although due to the bipolar nature of the geomagnetic field,  the events, such as the solar coronal mass ejections (CMEs) and associated solar wind enhancement, usually disturb the polar ionosphere only, the UV and X-rays from the solar flares produce sudden ionospheric disturbances (SIDs) in low-mid-latitude part of the earth's ionosphere. Such ionospheric disturbances are studied with the help of the influence they exert on radio waves propagating through earth-ionosphere waveguide. For the lower part of the ionosphere, called the D region, prominent modification in electron-ion density during solar flares can be observed via deviation in earth bound Very Low Frequency (VLF) wave signal from its ambient diurnal profile. In earlier work, successful model of the deviation in VLF amplitude due to different classes of solar flares was formulated. There, calculation of rate of ionization with Monte Carlo simulation and ion-chemistry evaluation of plasma density enhancement followed by a radio propagation simulation was used. Presently, we attempt to numerically reconstruct the modulation in VLF signal from its diurnal pattern produced by multiple solar flares occurring over a single day. Successful reconstruction of the VLF signal modulation for such a complex flaring scenario points to the accuracy of our understanding of the ionization effect due to solar activity on the lower ionosphere, and strengthen our claim to use earth's ionosphere as a high energy space transient event detector.
\end{abstract}

\begin{keyword}
\KWD Lower ionosphere \sep D region \sep Solar flares \sep VLF reconstruction \sep Monte Carlo simulation \sep Ion-chemistry model 
\end{keyword}

\end{frontmatter}


\section{Introduction}
\label{sec1}

\noindent Solar flares are sudden outbursts of enormous amount of energy stored in the Sun's magnetic field in the form of high energy radiation, mostly X-rays and extreme ultraviolet (EUV) photons. During solar flares highly energetic photons up to a few hundreds of keV come out from the burst region. Earth directed release of energy from such flares causes those energetic photons to primarily interact with the earth's upper atmosphere. The radiation penetrates most of the upper atmosphere ionizing the neutral molecules and depositing energy across the different layers. The lower energy X-rays ($\sim$ 2 - 12 keV) ionize the lower part of the ionosphere ($\sim$ 60 - 100 km above the earth surface), also called the D layer. The enhanced EUV also contributes in the enhancement of ionization. The extra ionization modulates the ionosphere by sudden increase of plasma density and consequently affecting radio wave propagation. The overall phenomena of such transient nature is termed as the sudden ionospheric disturbances (SIDs)\citep{Dellinger37, Mitra74}. \\

\noindent Depending on the X-ray brightness or amount of flux (measured in $Watt.m^{-2}$) in the wavelength range of 1 - 8\AA, solar flares  are categorized into different classes. The flares of C, M and X classes produce progressively stronger effects on the ionosphere through ionization. However, higher the class of the flare lower is the rate of occurrence. For example, X-class solar flares are much rarer than the C-class ones. The frequency of occurrence also vary with the phase of a solar cycle. During the peak of any solar cycle there may be multiple M-class solar flares in a single day and more than one X-class flare in a week. During the sudden onset of a larger flare the ionospheric plasma (electron-ion) density shoots up very quickly. After the peak of a flare is reached as the X-ray/EUV brightness starts decreasing, the plasma density of the lower ionosphere tends to go down as well but at a smaller rate than that of the flare intensity. The electron-ion recombination and other reaction processes at any height of the lower ionosphere are not instantaneous. The rates of those processes depend on the density of free electrons, ions, atoms and molecules present in the layer of the ionosphere and hence on the altitude within the layer. This causes delayed return of the different layers of the D-layer ionosphere to their respective original conditions. \\

\begin{figure}
  \centering
  \includegraphics[scale=0.33]{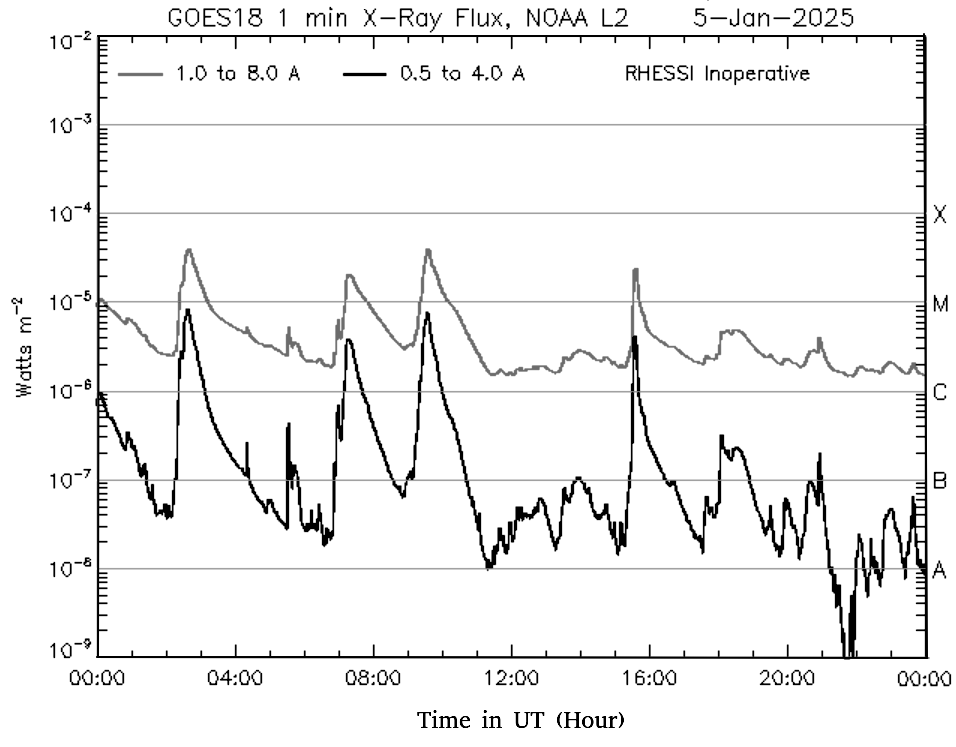}
  \caption{X ray fluxes in two different wavelength ranges, binned for one minute duration, for the day of $5^{th}$ of January 2025 showing multiple solar flares, mostly M-class ones, obtained from the GOES-18 data and plotted against proper setting of date and time at \url{http://sprg.ssl.berkeley.edu/~tohban/browser/}}
  \label{goes18}
\end{figure}

\noindent Very low frequency (VLF) radio waves (frequency $\sim$ 3 to 30 kHz) have been extensively used to study lower ionospheric disturbances by solar flares  \citep[and so forth]{Burgess67, Thomson2001, McRae04, zigman7, Han10, Kolarski11, palit13, basak13, singh2014, palit15, Bouderba16, Kumar18, Bekker21, Gu23, Ryakhovsky24}. The component of VLF wave  emitted from a transmitter situated on earth and reflected from the lower ionosphere carry the information on modification of the plasma density and effective reflection height during these ionizing phenomena originating from the Sun. During a flare induced SID the VLF signal can go through sudden increase or decrease depending on many factors including VLF wave frequency, emitted power, path of propagation and superposition of ground and sky (reflected from lower ionosphere) waves etc. \\

\noindent Numerous research using different approaches have been carried out to analyze such modulation in VLF due to solar flare induced SIDs. In one approach, VLF data are used to find the flare induced changes of ionospheric parameters, such as, effective reflection height ($ h^{\prime}$) and the sharpness factor ($\beta$) relating to slope of electron density variation with height. These derived parameters when combined with Waits' exponential formula \citep{wait64}, gives approximate change in lower ionospheric electron density \citep{zigman7, basak13, singh2014}. Another approach, which can be considered to be the forward reconstruction of the signal, starts from the in-situ observation of the flare photon (X-rays and EUV) emission and calculates the rate of ionization at different layers of the lower ionosphere as those photons pass through them. Then the electron-ion density produced are used in an ion-chemistry model to obtain equilibrium plasma density during the progression of the flare and ultimately a VLF radio wave propagation code is employed to find the probable VLF amplitude modulation \citep{palit13, Bekker21, Gu23}. This is then compared with observation. \\

\noindent \cite{palit13} successfully reconstructed the VLF amplitude modulation due to a M-class and a X-class solar flare using the latter approach. In their work GEANT4 Monte Carlo simulation package \citep{Agostinelli2003} has been used to calculate the ionization at different layers of the earth's lower ionosphere at a number of different points along the VLF propagation path between the transmitter and the receiver, and a very simplistic ion-chemistry model, namely, GPI model \citep{glukhov92} has been used to calculate the residual electron density values as a function of height during whole period of time when the flare occur. Ultimately, the Long Wavelength Propagation Capability (LWPC) code \citep{ferguson98} was used to calculate the change in VLF amplitude, which reproduced the relative change in VLF amplitude during the flares satisfactorily. The same model, adapted for much lower altitude (down to $\sim$ 30 km) has also been used to reconstruct the modification in VLF amplitude during a Soft gamma repeater or SGR \citep{palit18a,palit18b}. \\

\noindent Dealing with lower ionospheric response for a strong and isolated solar flare which is not preceded or followed by other such flares of comparable intensities is relatively straight forward and less complicated to be reconstructed with in situ simulation processes as mentioned above. However, for multiple solar flares occurring within a relatively short interval, the situation becomes more complex due to the following reasons: \\

\noindent During a solar flare, as the flare progresses, the change in plasma density in the lower ionosphere follows the flare intensity profile and hence the ionization rates with a delay imposed by recombination processes between the charged and the neutral particle components. The recombination processes are not instantaneous and take some time, depending on various reaction rates between the neutral and the ion components at various altitudes. As a consequence the enhancement in plasma density is delayed by certain amount at the onset of the flare as well as the enhanced plasma density takes longer time to return back to their original levels than the time taken by the flare to  subside. This implies that the enhanced plasma density values  in the lower ionospheric levels persist longer than the duration of the flare itself. In addition to this, the delayed recombination process also manifest in a time difference between peak of the flare and that of the response in VLF. This phenomenon termed as the VLF peak time delay has been explored by numerous workers. We suggest the reader the paper \cite{palit15} for a comprehensive review as well as the intuitive reasoning of peak time delay  demonstrated using in-situ ion-chemistry simulation and \cite{basak13} for quantitative investigation and correlation with flare intensity, class and solar zenith angle. The first paper computed the effect of recombination processes in producing delay in plasma density at different heights as well as its influence on VLF evolution for different classes of flares. In case of flares occurring in close succession the recombination delay causes the increased plasma density in the lower ionosphere due to a strong flare to remain in an enhanced level when the photons from the next flare start to reach the ionosphere, leading to an overlapping response in the D-region. This indicates the initial plasma density values accounted for the ion-chemistry model as part of the reconstruction simulation of the lower ionospheric modulation by a flare should be different from ambient values if there has been a flare preceding it and can not just be incorporated from source like IRI model \citep{rawer78} as was done in \citep{palit13}. \\

\noindent Secondly, consecutive large flares occur over several hours during which the zenith angle of the Sun undergoes significant change. In this scenario the ionization by ambient EUV becomes important, as throughout the daytime the EUV ionization rate varies with the zenith angle variation as described by Chapman formula \citep{chapman31}. This leads to a gradual variation of the lower ionospheric electron density and hence the ambient VLF amplitude throughout the day peaking during the noon. As a result the base plasma density encountered by the electrons and ions produced by X-ray photons received from different solar flares at different times of the day are not the same. Those from the flares occurring during the noon for certain position on earth naturally interact with enhanced plasma density at the lower ionosphere and during the morning and afternoon interact with lower plasma density. Therefore, the responses in the lower ionosphere in terms of enhancing plasma density incurred by two flares with identical X-ray brightness but occurring at different times of the day should be different. To account for this, while we are modeling multiple flares distributed throughout the day, we must incorporate the diurnal variation of EUV ionization in our simulation alongside the X-ray ionization mechanism. \\

\begin{figure}[h]
  \centering
  \includegraphics[scale=0.33]{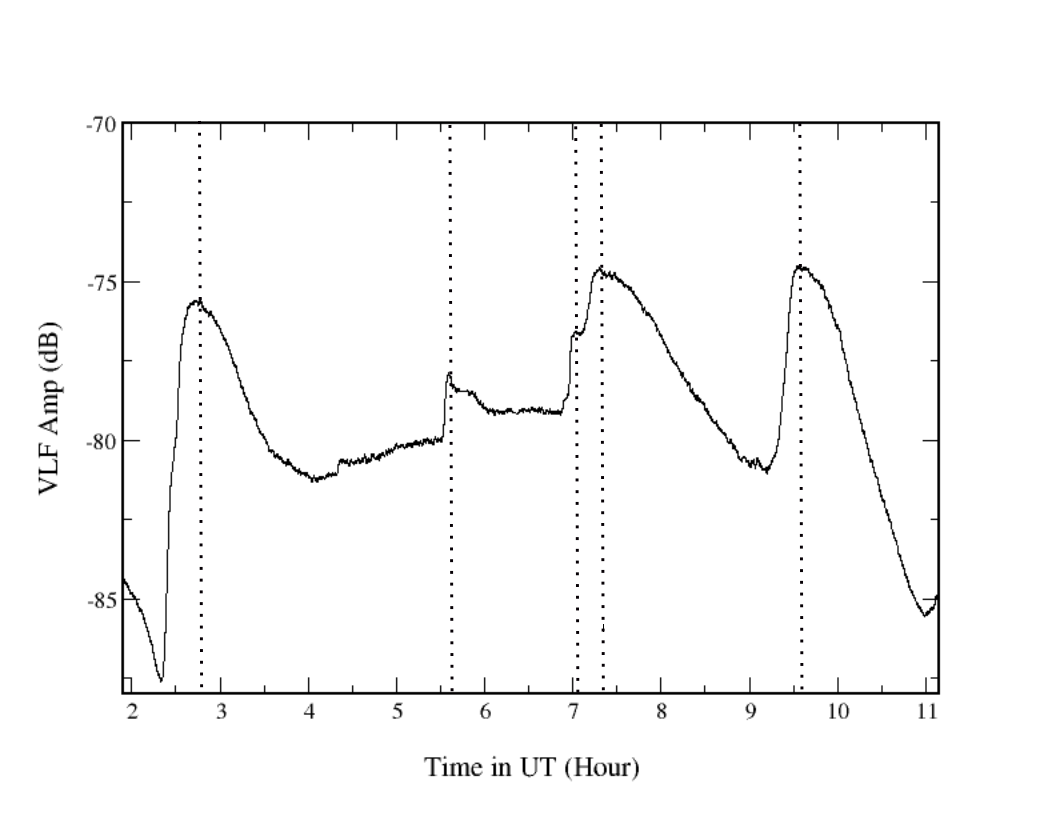}
  \caption{VLF amplitude for the whole day-time of $5^{th}$ of January 2025 transmitted from VTX transmitter (19.8 kHz) and received at ICSP receiver Kolkata is shown in the Figure. The dotted vertical lines correspond to peaks in VLF amplitude in response to the peaks of the transients or flares of diferrent brightness in solar soft X-ray light curve.}
  \label{amp}
\end{figure}

\noindent In this paper we attempt an in-situ modeling of the VLF amplitude modulation observed in a single day due to several solar flares in the $25^{th}$ solar cycle.  There are three prominent M-class flares and few smaller peaks in X-ray brightness distributed throughout the whole day-time. There is a small peak at the time when the brightness of the first M-class flare is decreasing and almost subsides. Another small peak occurs just during the time of commencing of the second M-class flare. The effect of all of these large and small brighness peaks have been clearly observed in the VLF data presented in this paper. We attempt to compute ab initio and reconstruct these flare modulations using in-situ calculation. The organization of the paper is as follows. In Section 2, we describe the observation of the VLF data as well as those required for simulation purpose. In Section 3, we briefly describe the method adopted for the reconstruction simulation. In the next Section, i.e., Section 4, we describe our result and in the last Section we discuss on the outcome, limitations and  future direction of work. \\

\section{Observation and data accumulation for simulation}
\label{obs}

\begin{figure*}[t]
  \centering
  \includegraphics[scale=0.23]{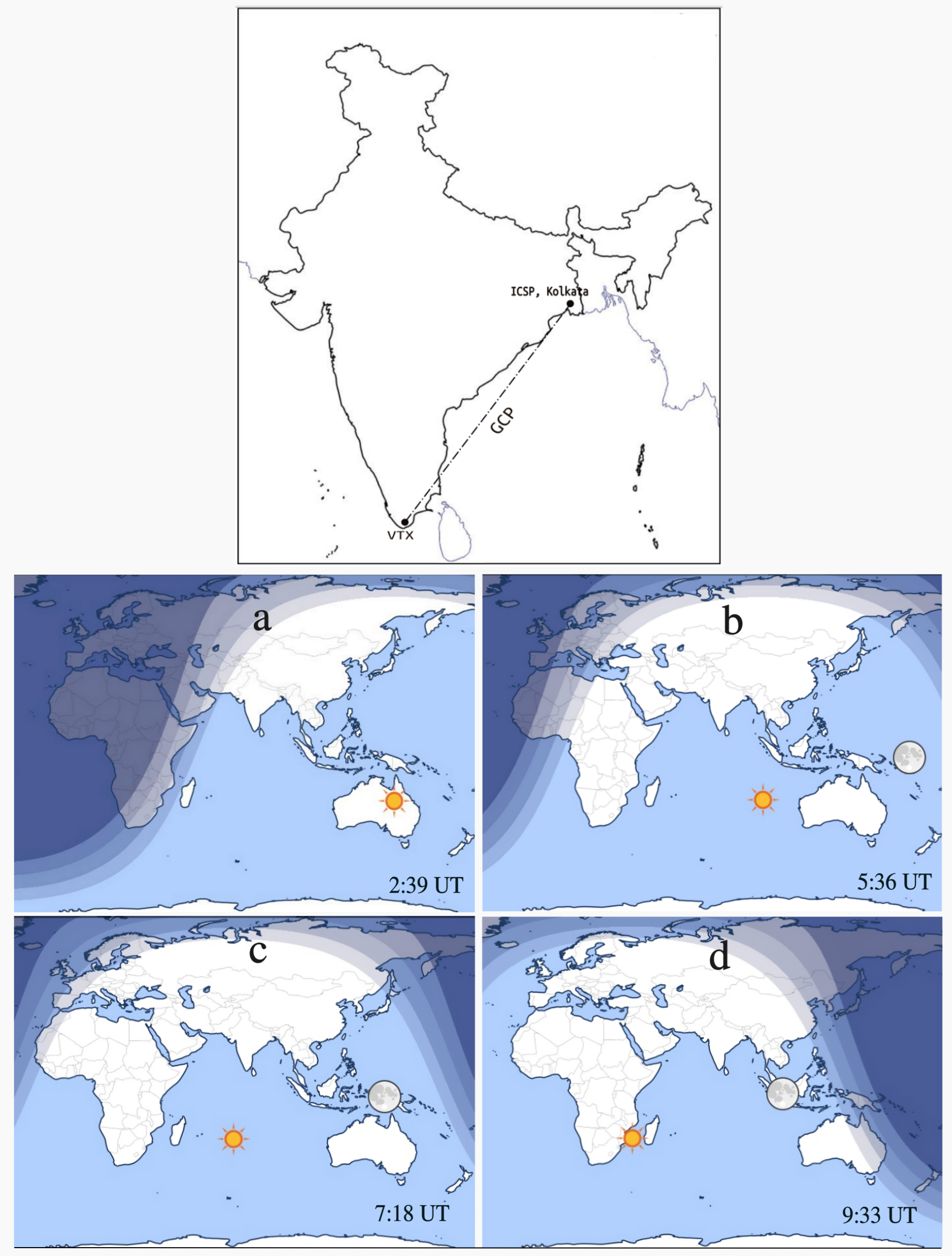}
  \caption{The Top panel shows the VLF great circular path (GCP) between the transmitter VTX and the receiving station at ICSP, Kolkata. The Figures in the Bottom panel show the subsolar points during the peaks of the first three M-class solar flares (a,c,d) and the C-class flare (b) at 5.5 hr (UT) the effects of which on the VLF amplitude are shown by verticle dashed lines in Figure~\ref{amp}. The Bottom Figures are obtained using Day and Night World Map of https://www.timeanddate.com. At the terminators the lightest to darkest twilight shadows represent Civil, Nautical and Astronomical Twilight respectively. }
  \label{path}
\end{figure*}

\noindent With the nearing of the peak of the $25^{th}$ solar cycle, which is predicted to occur in $\sim$ July, 2025, there have been frequent large (M and X class) solar flares. Often more than one large flares are occurring in a single day. We choose the data of one such day, namely, $5^{th}$ of January 2025, in which four M-class flares and few smaller peaks in the X-ray brightness occurred in the whole day as shown in the Figure~\ref{goes18}. The Figure is obtained from the RHESSI Quicklook browser interface by NASA and shows GOES-18 1-minute X ray fluxes in two different wavelength ranges for the day. The data with the range of 1 -- 8$\AA$ (1.55 -- 12.40 keV) is shown in light and the data  with the range of 0.5 -- 4$\AA$ (3.1 -- 24.79 keV) is shown in deep black colored lines. \\

\noindent We can see that for the whole day the X-ray brightness persists above the level corresponding to C-class flare. For near solar maximum condition in the 11 year solar cycle this is quite usual. The first strong flare, which peaked around 2:39 AM (UT) is of M4.1 class, the second flare peaking around 7:18 AM (UT) is a M2.0 class one. The third flare which is also a M4.1 class peaked around 9:33 AM (UT) and the 4th M2.3 class flare peaked around 3:36 PM (UT). Other than these large flares there are few smaller bumps in the X-ray light curves within the range of C-class brightness. Out of all these flares the first three M-class ones and two smaller peaks in X-ray brightness occur in the daytime at our receiver location, namely ICSP, Kolkata (Lat: $22.49^\circ N$, Lon: $88.40^\circ E$). We could detect these flare signals in VTX ($8.43^\circ N $, $ 77.73^\circ E$, frequency 18.2 kHz) and NWC ($21.8^\circ S $, $ 114.15^\circ E$, frequency 19.8 kHz) transmitter data, but as the NWC data is very noisy we use the VTX data only for our investigation. The approximate sunrise and sunset terminator times for the diurnal variation of VLF at that point are taken to be 1.4 UT during the morning and 11.1 UT during evening respectively. In the Figure~\ref{goes18} we see there are 3 large peaks corresponding to the first three M-class flares and one small peak at around 5.5 hr (UT) and another slightly larger peak at around 6.9 hr (UT) just during the onset of the second M-class flare appearing in the time period between the two terminators. In the VLF amplitude data in Figure~\ref{amp} we can clearly identify the effect of the larger and smaller peaks, where the peaks in VLF amplitude modulated by the respective flare modulations are marked with dotted vertical lines. We find that the times of occurrence of all the VLF amplitude peaks are slightly shifted to the right with respect to the corresponding 
flare X-ray brightness peaks. \\

\noindent In Figure~\ref{path} we show the great circular path (GCP) of VLF propagation between the VTX transmitter and the ICSP, Kolkata VLF receiver in the Top panel. In the Bottom panel we show the subsolar points and the sunlit and shadow regions of the part of the world during the peaks of the first 3 M-class solar flars and the C-class flare at 5.5 hr (UT). These give us an idea of how the solar zenith angle and corresponding strength of solar radiation varies over time during the occurence of the flares througout the day. This effect has been incorporated in the simulation of ionization described in next Section.\\

\begin{figure}[h]
  \centering
  \includegraphics[scale=0.33]{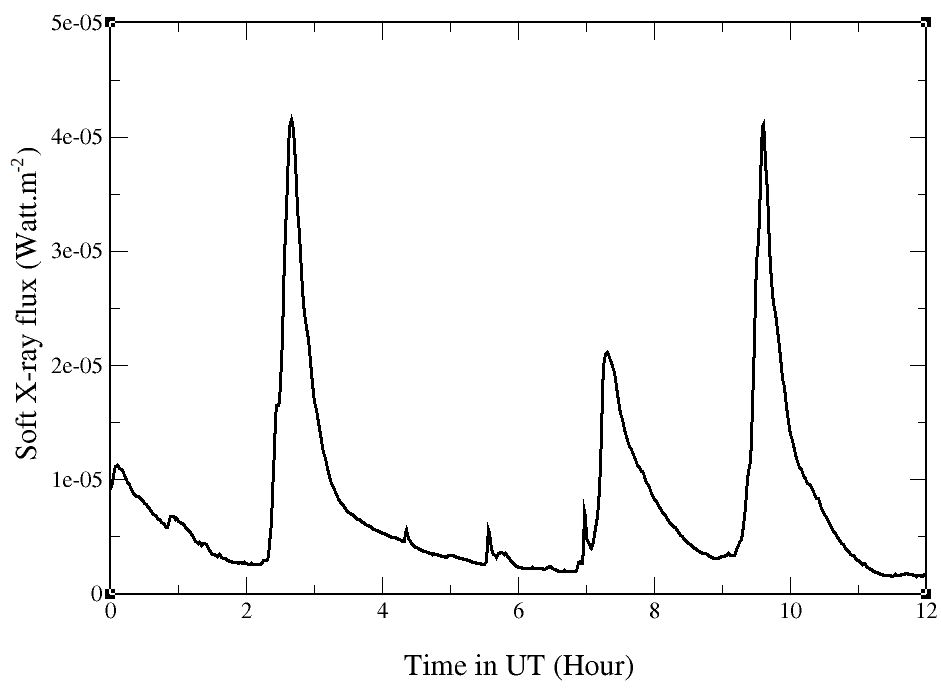}
  \caption{Soft X-ray flux from 1–8 $\AA$ (1.55 – 12.40 keV) obtained from GOES satellite data of the first 12 hours (UT) of $5^{th}$ of January 2025}
  \label{xraylc}
\end{figure}

\begin{figure}
  \centering
  \includegraphics[scale=0.33]{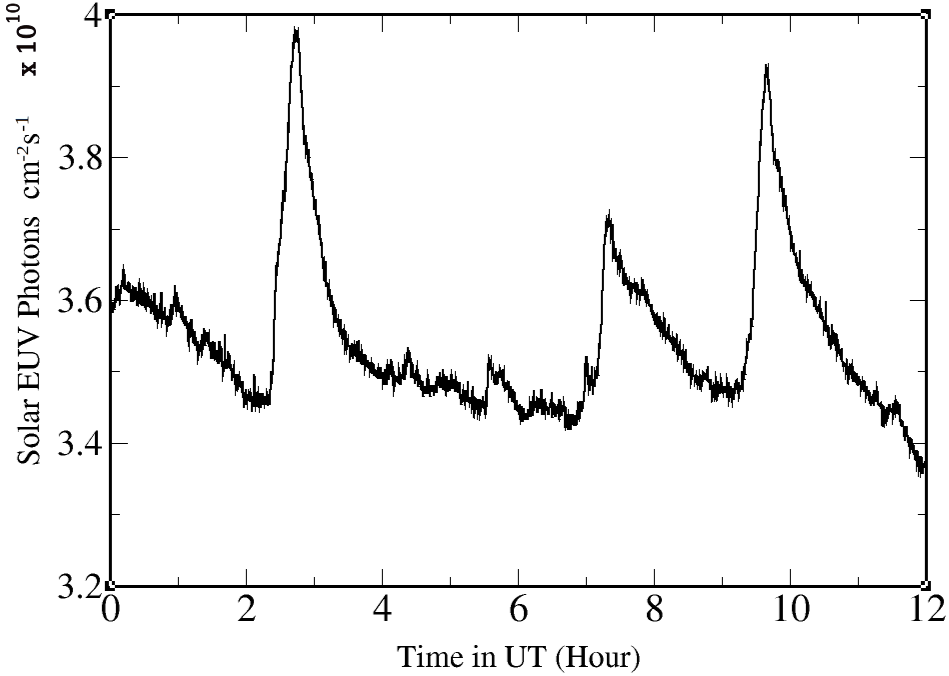}
  \caption{UV photon flux in unit of $photon.cm^{-2}s^{-1}$ in the range of 0.1 - 50 nm obtained from SOHO/SEM for the the first 12 hours (UT) of $5^{th}$ of January 2025}
  \label{uvlc}
\end{figure}

\noindent We found the GOES Longwave (1–8 $\AA$) flux ($watt. m^2$) from Space Weather Data Portal and plotted as function of time (UT) in Figure~\ref{xraylc} for the first 12 hours of the day. In Figure~\ref{uvlc} we plotted the EUV (0.1 - 50 nm) photon flux of the first 12 hours of the same day obtained from SOHO/SEM. From Figure~\ref{xraylc} we find that there is almost 10 times increase in X-ray flux during the larger flares from the ambient values, whereas Figure~\ref{uvlc} shows for EUV the increase is more or less $\sim 0.1$ times the ambient values. From the comparision of relative increase in flux values of the two components during flares, it becomes obvious that the increased X-ray brightness, or more specifically soft-X ray emission during the flares takes the dominant role in modulation of lower ionospheric behaviors and hence VLF signal amplitude.\\

\section{Methodology}

\begin{figure*}
  \centering
  \includegraphics[scale=0.37]{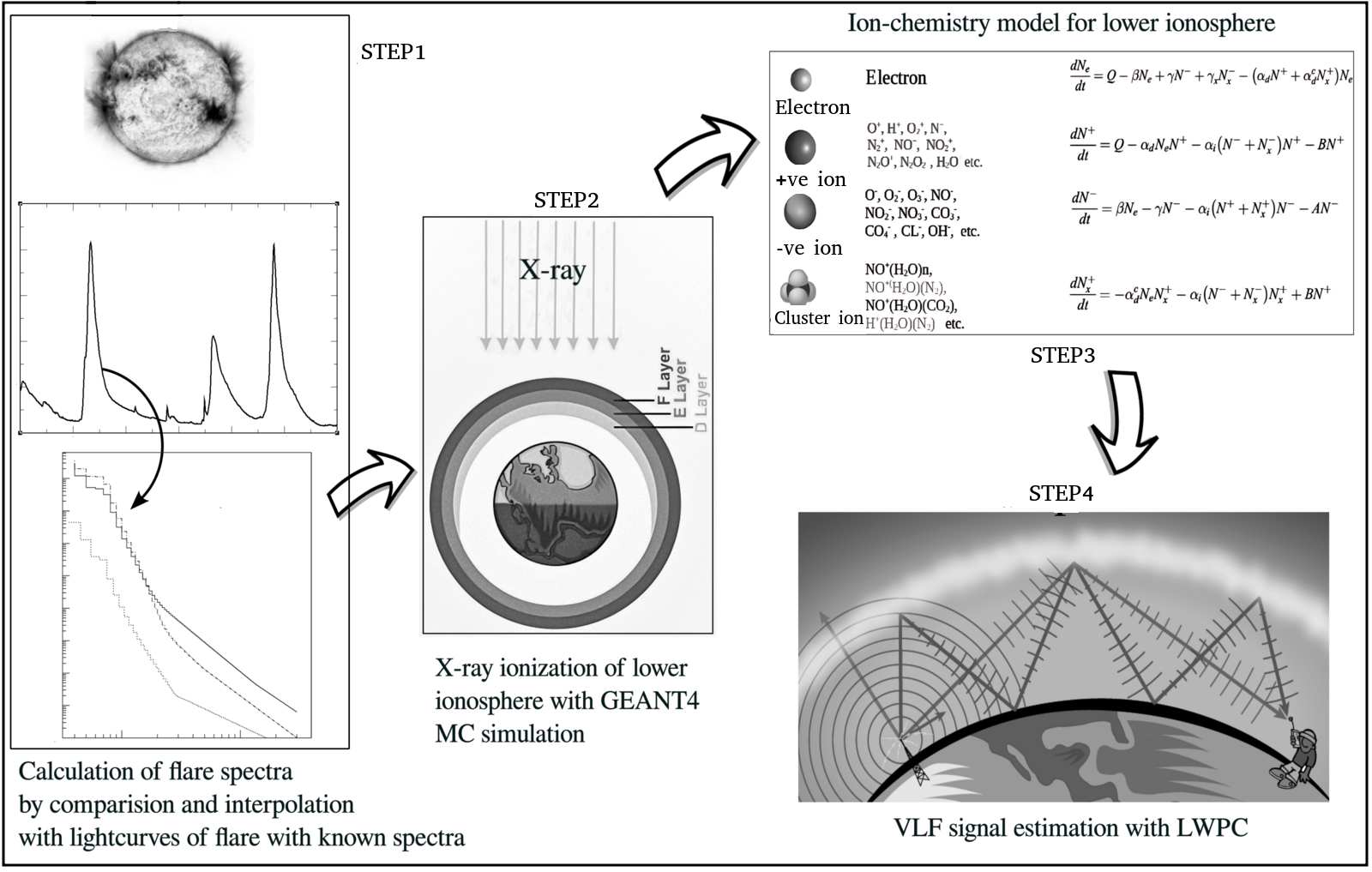}
  \caption{A diagram depicting the model steps employed in the work}
  \label{model}
\end{figure*}

\noindent To reconstruct the observed modulation in diurnal variation of VLF signal amplitude by strong solar flares on the day we start with the calculation of number of high energy photons received by different layers of the lower ionosphere in form of both X-rays and EUV from those flares. During the strong M-class flares the soft X-ray brightness increases up to $\sim$ 10 times or more than the ambient condition as seen in Figure~\ref{xraylc}. The EUV photons are always present irrespective of the presence of solar flares, but during the flares there is around 10 \% increase in EUV as found from Figure~\ref{uvlc}. Both the increment in X-rays and EUV fluxes should cause extra ionization in the earth's lower ionosphere resulting sudden enhancement in its plasma density. This in terms should affect the signal modulation of VLF radio wave component reflected from the part of the ionosphere. \\

\begin{figure*}[t]
  \centering
  \includegraphics[scale=0.3]{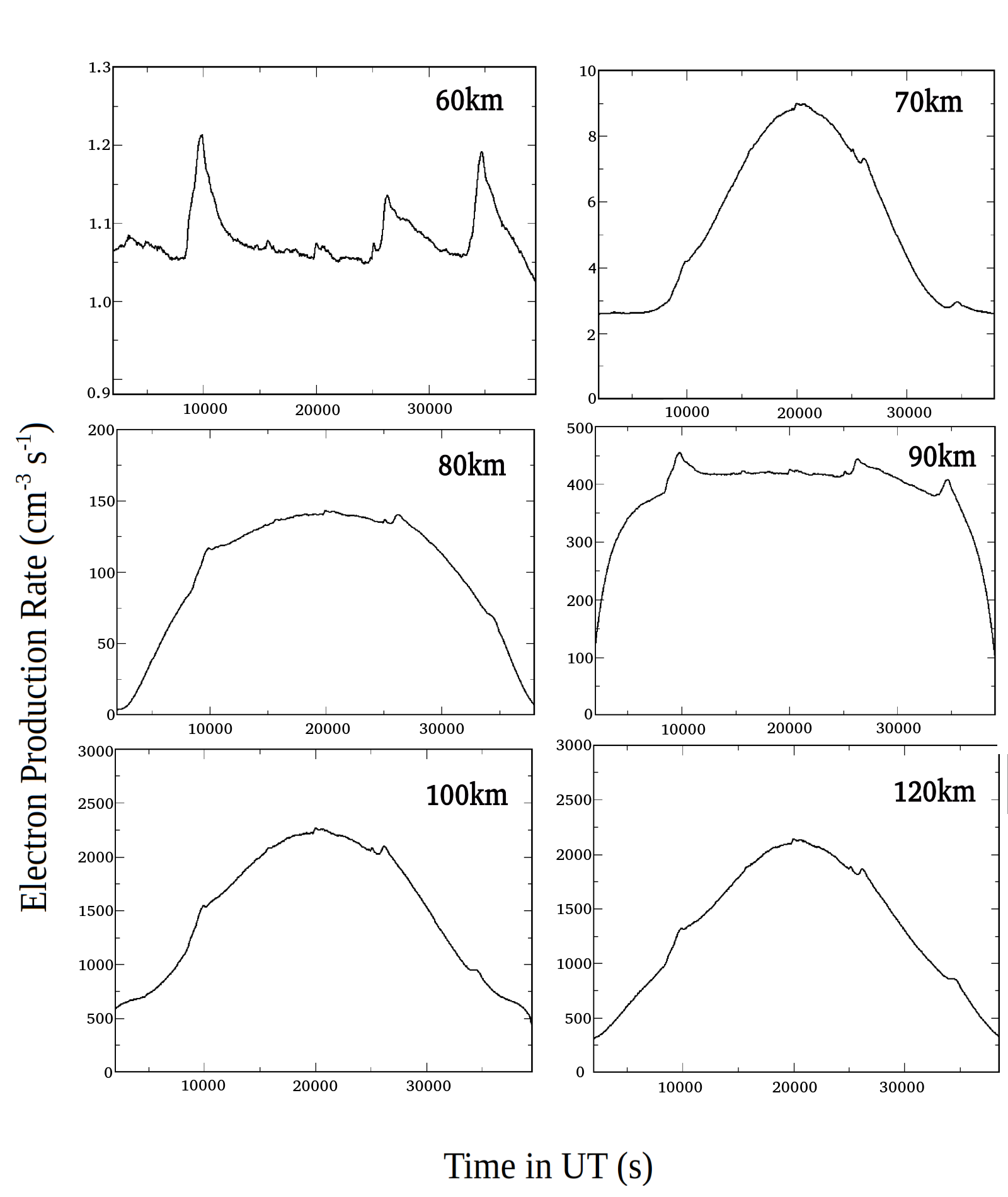}
  \caption{Computed ionization (or free electron production) rates ($cm^{-3}s^{-1}$) due to EUV photons as function of time at various heights for the day of VLF reconstruction}
  \label{eproduv}
\end{figure*}

\noindent The whole process of simulation comprises of four main stages (see Figure~\ref{model}). First we find the ligthcurves and approximate spectra of the solar flares both for X-ray and UV rays to calculate ionization rates. Then using those spectra and lightcurves we estimate the altitude variation of electron-ion production rates by the ionization of enhanced EUV and X-ray photons in the lower ionosphere ($\sim$ 60 - 140 km). In the next step we combine the calculated ionization rates and use it in a simplistic ion-chemistry model to find residual electron density at any point of time determined by both production and loss processes of the free electrons and ions. Finally, LWPC code is used to find the modulation in the VLF amplitude transmitted by the aforementioned transmitter (VTX) and observed at the receiver point at ICSP Kolkata.\\

\subsection{Calculation of Ionization}
\noindent It is obvious that the relative contribution of enhanced EUV in electron-ion production and hence increasing the lower ionospheric electron-ion density during such flares should be minimal as compared to that by increased soft X-rays. As discussed in the Introduction Section the diurnal variation of ambient plasma density varies in response to zenith angle variation of EUV ionization throughout the day. This should impact the response of X-rays from flares occurring at different times of the day. So, unlike what was done in earlier work \citep{palit13}, we incorporate EUV ionization here as well. We have examined both the contribution of EUV and X-ray photons in electron-ion production rate as function of height and time, combine them and input in our ion-chemistry code to calculate the enhancement in plasma density. \\

\subsubsection{Ionization due to EUV photons}
\noindent The altitude ($h$) variation of solar EUV ionization rate ($q_{uv}$) as a function of time ($t$) can be calculated using a modified Chapman’s formula \citep{chapman31, mitra51} of ionization rates. As described by \cite{rees89},

\begin{eqnarray}
\begin{split}
q_{uv}(h,t)=\sum_{j}\int I_0(\nu,t) e^{\sum_{k}\sigma_k(\nu) \int_{h}^{\infty} n_k C_h(h,\psi)dh} \\
\times \eta_j(\nu) \sigma_j(\nu)n_j(h)d\nu,
\end{split}
\end{eqnarray}

\noindent where, $I_0(\nu,t)d\nu$ is the  solar flux or irradiance in the frequency range $\nu$ to $\nu+d\nu$ at the top of the atmosphere (TOA). $\sigma_j(\nu)$ is the absorption cross section for the $j^{th}$ neutral component of air at that frequency interval, and the $\eta_j(\nu)$ and  $n_j(h)$ are the photo ionization efficiency and concentration for the $j^{th}$ component respectively. The grazing incident function, $C_h(h,\psi)$, which is a function of solar zenith angle  $\psi$ and altitude, plays a significant role  and  for a certain altitude $h_a$ is given by

\begin{eqnarray}
\begin{split}
\int_{h_a}^{\infty} n_jC_h(h,\psi)dh=\int_{h_a}^{\infty} n_j[1-(\frac{R_e+h_a}{R_e+h}) ^2sin^2(\psi)]\\
\end{split}
\end{eqnarray}

\noindent for $\psi <90^\circ$ \\
and,

\begin{eqnarray}
\begin{split}
=2 \int_{h_{min}}^{\infty} n_j[1-(\frac{R_e+h_{min}}{R_e+h}) ^2]dh \\
 \quad  -\int_{h_{min}}^{\infty} n_j[1-(\frac{R_e+h_{min}}{R_e+h}) ^2sin^2(\psi)]^\frac{1}{2} dh
\end{split}
\end{eqnarray}

\noindent for $\psi >90^\circ $. \\

\noindent Here, $R_e$ is the earth's average radius and $h_{min}=cos(\psi-90^\circ)(h_a+R_e)-R_e$ is defined as the minimum height of photon path.\\

\noindent The photo-ionization efficiency, absorption cross section etc. have been presented in various articles such as \citet{ohsh66} for $NO$ and $Ar$,  \citet{torr79} for $N_2$, $O_2$ and $O$, \citet{mcew75} for $CO_2$. Those values have been collected and tabulated in an earlier article \citep{chak16}. Along with those the reference EUV spectrum of the Sun for the value of F10.7 = 68, from \citet{torr79} are also tabulated there. EUV flux is dominated by Lyman-alpha line, and irrespective of the occurrence of flares the respective time variation of the Lyman alpha and UV continuum are very similar implying that the ratio of energy emitted in different energy bands remain almost constant. We use this fact to normalize the reference spectrum with the  0.1 - 50 nm flux (averaged over 15 sec) data, obtained from the Solar EUV monitor (SEM) of Solar heliospheric observatory (SOHO)\footnote{\url{https://lasp.colorado.edu/eve/data_access/eve_data/lasp_soho_sem_data/long/15_sec_avg}} throughout the simulation period. The normalization is done by multiplying the whole reference spectrum with the ratio of instantaneous 15 s average EUV fluxes in the 0.1 - 50 nm range (Figure~\ref{uvlc}) to the cumulative flux of the tabulated reference spectrum in the same wavelength range. It gives us the approximate 15 sec average EUV spectra for the whole time of simulation.\\

\noindent In Figure~\ref{eproduv} we plot the rate of electron-ion production as function of time at different altitudes due to solar EUV as calculated above for the whole day time. We see signature of flare profile superposed on diurnal variation of ionization rates at different altitudes. The increase is not very significant for any of the flares considering most of the electron-ion produced at certain point of time at any height prompty recombine through chemical reactions. The maximum relative increase in ionization from ambient values, $\sim 10$ \%, has been found during the peak of the first M-class flare at $\sim 60$ km altitude.  \\

\begin{figure}
  \centering
  \includegraphics[scale=0.28]{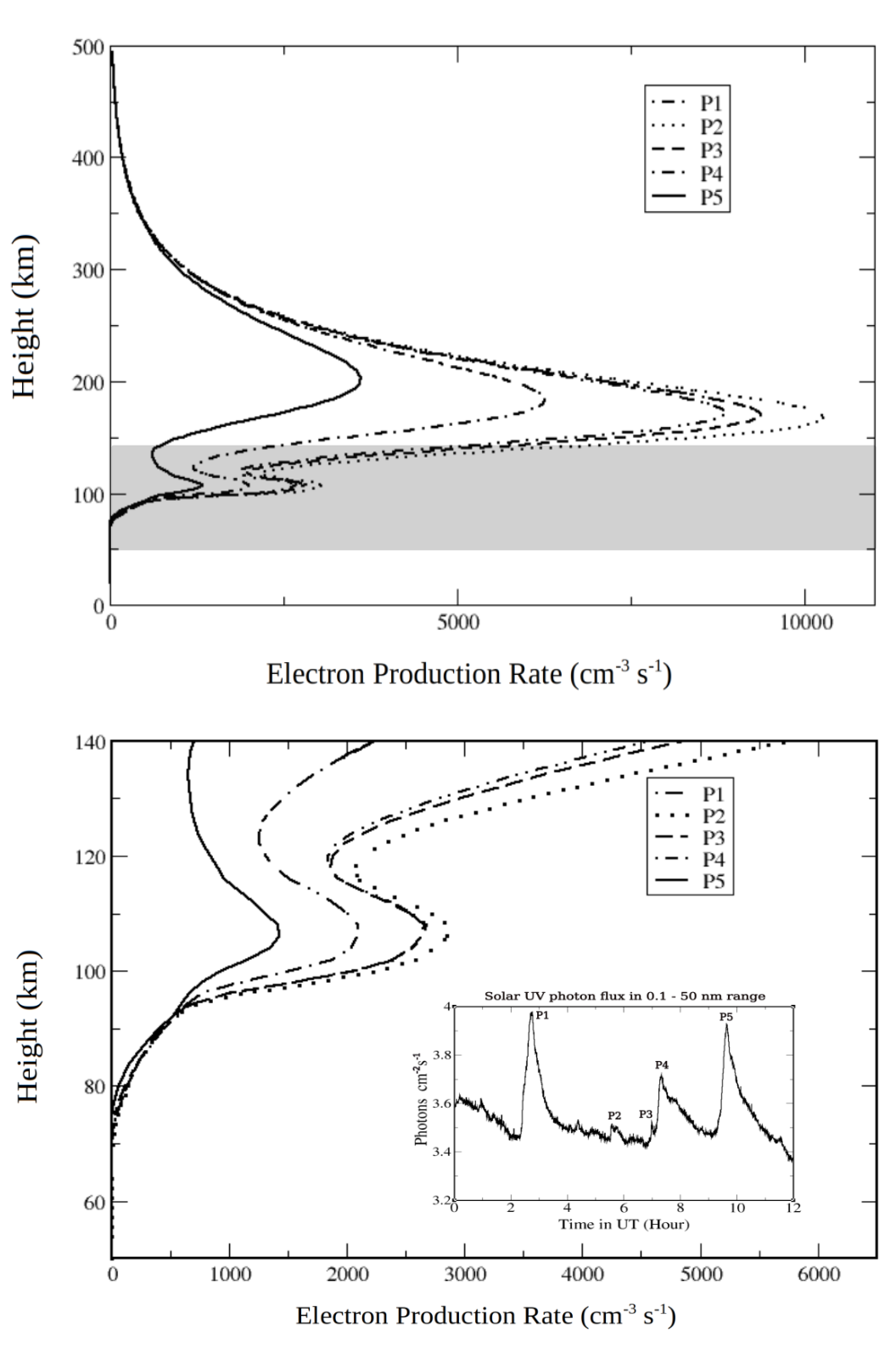}
  \caption{Ionization rate ($cm^{-3}s^{-1}$) due to EUV photon as function of altitude at different peaks of solar activity of the day is depicted. The Top panel presents the UV-induced ionization rates up to the altitudes where the production becomes negligible. The shaded region marks the lower ionospheric heights, which are displayed in greater detail in the Bottom panel}
  \label{eproduvht}
\end{figure}

\noindent In the Figure~\ref{eproduvht} we show the electron density production rates due to solar EUV as function of height during the peaks of five solar flares on that day those affected the VLF diurnal signal at Kolkata station. Obviously the ionization rate is a function of the solar EUV flux as well as the zenith angle. From the Figure it is evident that during the peaks of the two big M-class flares, one of which occurs during the morning and the other just before the evening, less ionization is produced than during the other smaller peak and the M-class flare occurring around the  middle of the day, when the sun was near the zenith. For example, the ionization rates at all the heights during P2 is much greater than those during P1, although P2 is tiny compared to the peak P1. Also, very similar and almost indistinguishable profiles of electron production rates as function of height during the two peaks, P3 and P4 are found, although the first peak (P3) is a very tiny one compared to the large M-class flare peak denoted by P4. These imply that, when it comes to EUV ionization by flares, it is the solar angular position or zenith angle that plays more significant role than the flare's intensity or class. 

\subsubsection{Flare X-ray Ionization}
\noindent To calculate the enhancement of ionization in the earth's lower ionosphere by the enhanced X-rays during the flares we use Monte Carlo simulation, which is  performed with GEANT4 Monte Carlo simulation software. For the details of the simulation set up and process we refer the readers to \citet{palit13}. The earth's atmosphere is set in the GEANT4 geometry module in the form of concentric layers of air with its properties adopted from the NASA MSIS90 model \citep{hedin91} with increasing thickness as we go up to around 500 km altitude. Monochromatic beams of X-ray photons from 1-100 keV are made to impinge on the top of the atmosphere from various incidence angles and their responses as electron production rates per unit volume at different heights (at 2 km interval) of the atmosphere are stored in the computer memory. At any particular instant of time the total ionization due to the X-rays from the flare in that height range can be found by multiplying the corresponding response (adjusted for the zenith angle corresponding to that time of the day) for monochromatic beams at that height with the number of photons in that particular flare spectral (energy) bin and then summing them up. For this we need flare spectra at a small interval ($\sim$ 1 sec) of time during the evolution of the flare.\\

\begin{figure}[h]
  \centering
  \includegraphics[scale=0.33]{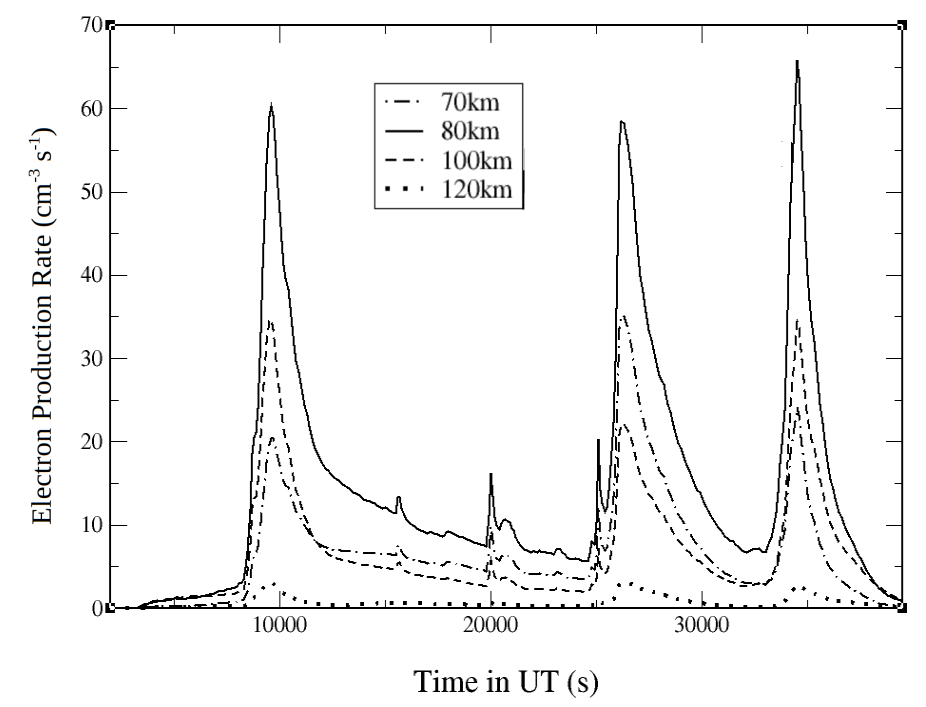}
  \caption{Computed ionization rate ($cm^{-3}s^{-1}$) due to X-ray photons as a function of time at various heights throughout the day}
  \label{xprod}
\end{figure}

\begin{figure}
  \centering
  \includegraphics[scale=0.33]{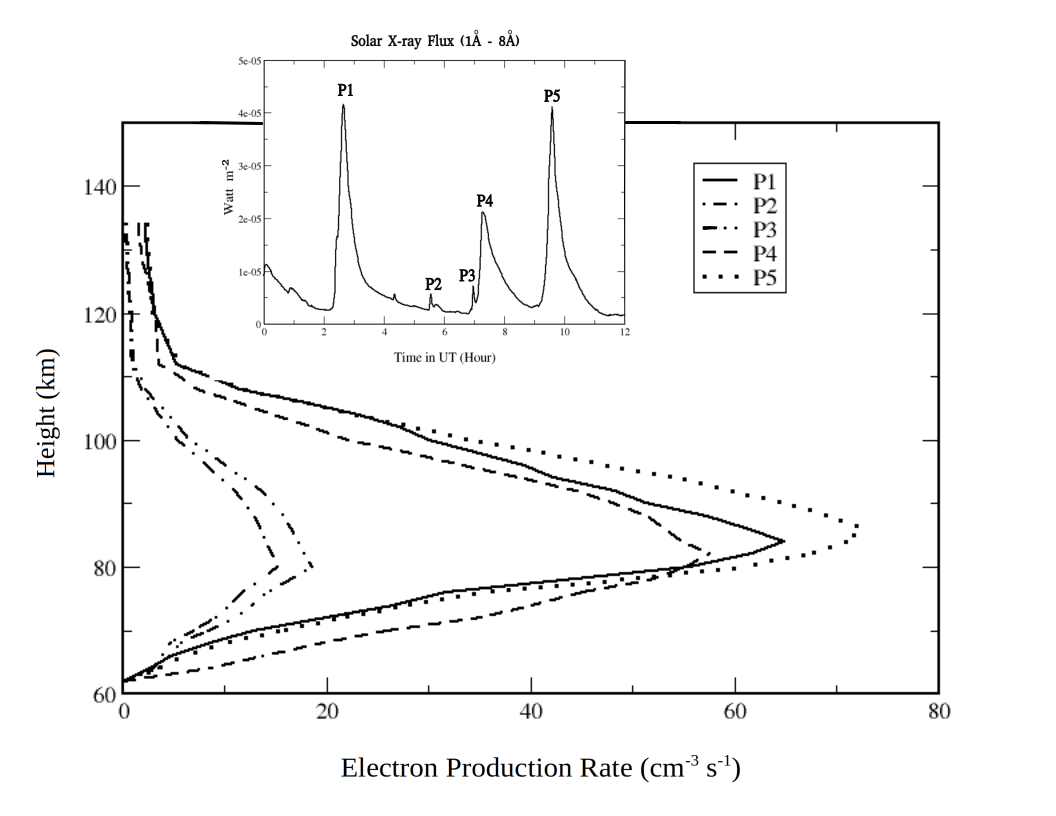}
  \caption{Ionization rate ($cm^{-3}s^{-1}$) due to X-ray photon as a function of altitude during different peaks of the solar flare soft X-ray brightness of the day}
  \label{xprodht}
\end{figure}

\noindent During an earlier work presented in \citet{palit13} the solar X-ray spectral data from the RHESSI satellite \citep{sui02} was used. There a few reference spectra during the ascent, peak and descent phases of the flares were used and bin-wise interpolation between them at each one sec interval was made to find the required spectra for each second during the flare evolution. This is required to calculate the ionization distribution in each second interval needed by the ion-chemistry code to be deployed in the next step. Though the approach is approximate and does not account for the change in spectral slope during the progression of the flare with utmost accuracy, it has been found to have produced very satisfactory result as described in \citet{palit13}. Now that the RHESSI satellite has ceased operation and no suitable alternative is currently available for obtaining solar flare X-ray spectra, we have made a further approximation. We assume that two solar flares belonging to the same class and exhibiting similar peak X-ray flux values follow a comparable photon energy distribution at any given stage of their evolution defined by identical X-ray flux levels. Based on this assumption, we normalize the reference spectrum of one flare to the flux level of another similar flare, allowing us to estimate its approximate spectrum at a specific point in time.\\

\noindent  In this work, we employ reference spectra from an M-class solar flare previously analyzed in \citet{palit13} to approximate the dynamic spectra for the M-class flares under current investigation. First, we identify reference points of those spectra in the light curve of the earlier flare and locate corresponding points in the light curve of the current data (Figure~\ref{xraylc}) that match those flux values. We then assume that the reference spectra correspond to those matched points in the present flare's light curve. Using this mapping, we perform spectral interpolation at one-second intervals for each spectral bin, thereby constructing the required spectra for input into the ion-chemistry simulation. We acknowledge that this approach introduces a level of scientific uncertainty and may lead to errors in the calculated electron-ion production rates, residual electron density, and consequently the VLF response. However, we proceed with this approximation under the expectation that the associated error is likely smaller than other inherent uncertainties in the model, such as those arising from intutive choices of chemical reaction rates \citep[see][]{glukhov92}, radio wave propagation, and wave-particle interaction calculations used to reconstruct the VLF signal.\\

\begin{figure*}
  \centering
  \includegraphics[scale=0.25]{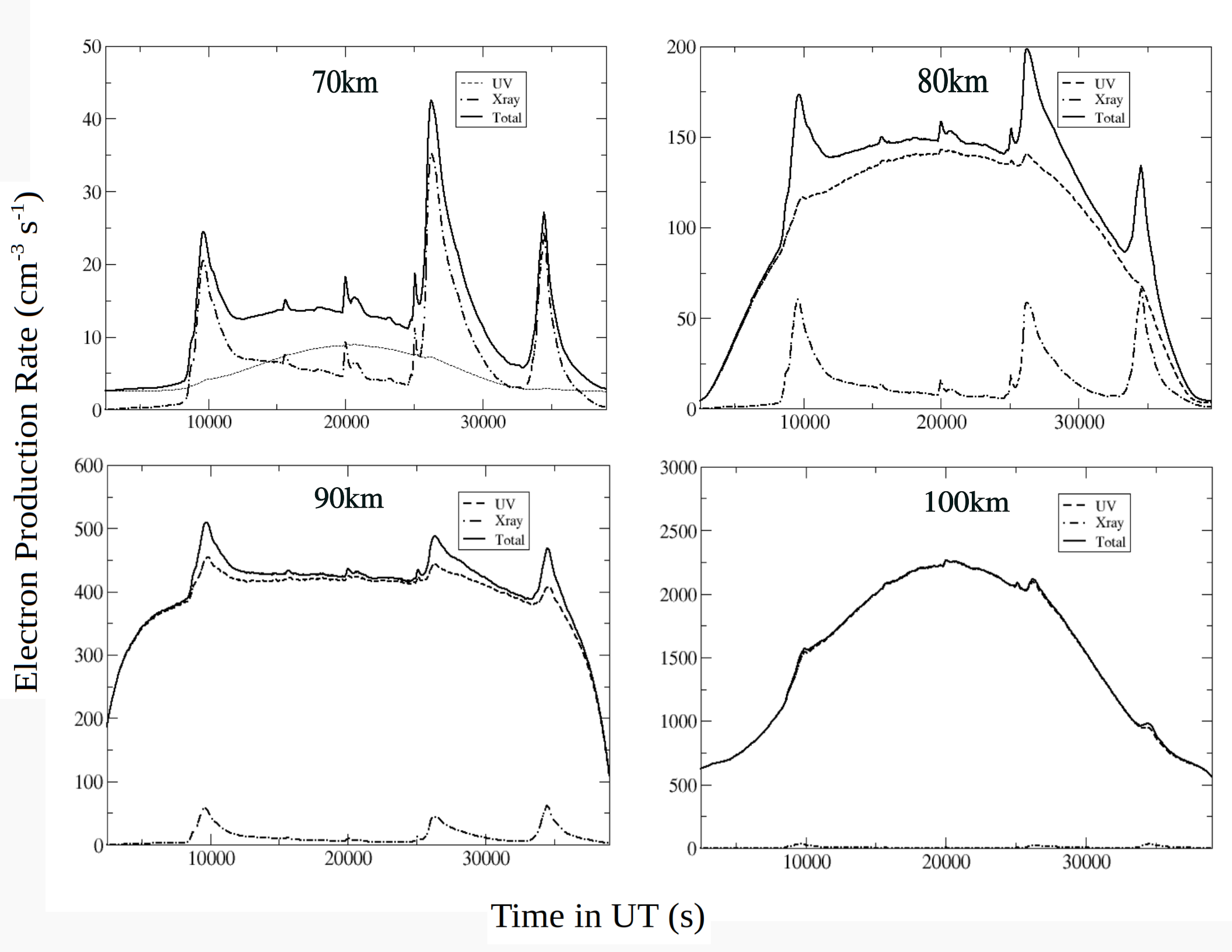}
  \caption{Simulated ionization rates by the X-rays, EUV and Total (X-ray + EUV) ionization rate as function of time at different altitudes: The total ionization rate is depicted by the solid line, whereas the ionization due to X-ray and EUV are represented by the dot-dashed line and dashed lines respectively.}
  \label{eprodtot}
\end{figure*}

\noindent The Figure~\ref{xprod} depicts the variation of ionization rates due to the X-rays over the day at different altitudes corresponding to lower ionosphere. From the Figure it is clear that the maximum ionization due to the soft X-ray photons occurs at the height around 80 km altitude, that can also be corroborated from Figure~\ref{xprodht}, which shows the ionization rates during the different peaks as function of height. Below $\sim$ 60 km and above  $\sim$ 110 km  the rate of ionization during the flares are very low. The Figure~\ref{xprod} also demonstrates that the variation of solar angular position throughout the day has negligible effect on X-ray ionization. This is not surprising as \citet{palit13} has previously shown that although the ionization rates at different heights peak when the Sun is at the zenith, those do not reduce much until the solar zenith angle reaches $\sim 70^\circ$, after which the rates decrease rapidly. In the present case the flares for most of their duration evolve when the solar zenith angle in the location of the great circular path (GCP) between the transmitter and the receiver remains within $\sim 70^\circ$. Also, unlike the EUV radiation the Sun does not emit much of X-ray photons other than times of occurence of flares,  so in between the flares there is almost negligible X-ray photons ionizing the lower ionosphere. Hence unlike the regular increase and decrease of ionization rates by ambient EUV photons depicted in Figure~\ref{eproduv} at most of the heights there are no such gradual variation seen in Figure~\ref{xprod}. The Figure~\ref{eprodtot} shows the time variation of ionization rates due to the diurnal variation of X-rays and EUV at four different altitudes and the combined ionization rates. These give us the clear understanding of relative contribution in ionization rates due to the X-rays and EUV from the flares. It can be seen that there are prominent increase in the ionization rates at around  70 - 90 km altitude during solar flares and most of this increase is coming from flare X-ray ionization. The increased ionization rates due to enhanced EUV during the flares are almost negligible compared to the other counterpart in this altitude range. However, above this range the ionization by enhanced X-ray is also scarce as depicted by the bottom-right picture of Figure~\ref{eprodtot} showing the ionization rate evolution at the altitude of 100 km.

\subsection{Ion-Chemistry Model}

\noindent The electrons and ions produced by the enhanced X-ray and EUV photons undergo various chemical interaction processes in the lower ionosphere, such as electron attachment with neutral atoms to form negative ions, electron detachment from negative ions, dissociative recombination-a process where a molecular ion recombines with an electron, resulting in the neutral molecule dissociating into fragments, recombination with positive cluster ions, ion-ion recombination process etc. Some of these processes contribute in overall increase of free electron density and some in decrease of it. We have adopted a simplified ion-chemistry model originally proposed by \citet{glukhov92}, called GPI model, that has been used previously in many of our works, such as \citet{palit13, palit15, palit18a, palit18b, chak16} and so forth. In this model all the charged particles are divided in four broad categories, namely free electrons, positive ions, negative ions and positive cluster ions.  Four coupled differential equations corresponding to each of the species are solved with $4^{th}$ order Runge–Kutta method to track their evaluation in the presence of free electron and ions produced by the solar X-rays and EUV photon ionization as calculated with GEANT4 Monte Carlo code and Chapman's formula collectively. The charge neutrality conditions are maintained throughout, i.e., the sum of positively charged ions and that of negatively charged ions and electrons are kept equal. The ionospheric neutral parameters like density and temperature are imported from the NASA MSIS 90 model and the initial values of the charged particles, namely electrons and ions are used from IRI model \citep{rawer78}. The details of the computation process and choice of the reaction coefficients used in the model can be found in \citet{palit13}. The successful execution of the ion-chemistry model provides us the electron density at each second throughout the day, both at the presence and absence of the flares as a function of height from 60 - 140 km altitude. \\

\subsection{VLF reconstruction with LWPC code}
\noindent  We use the LWPC code \citep{ferguson98} to compute the time variation in VLF signal amplitude throughout the local day time. RTABLE program of the LWPC
is executed, which gives the VLF amplitude and phase at any point of time, provided the altitude distribution of electron densities are given at few equidistant points along the GCP between the transmitter and receiver. We use four points, including the transmitter and the receiver points, which are at same spatial intervals in the GCP. For all the four points we calculate the electron density distribution as function of time at 2 km altitude interval from 60 to 140 km. Taking four such positions to incorporate electron density values is found to be sufficient and adding more points does not improve the model results. It also requires the height distribution of electron-neutral collision frequency, which plays an important role for the propagation and so-called reflection of VLF signals. The collision frequency profile between the electrons and neutrals that has been used in our calculation is obtained from \citet{kelly2009}, which is given by $\nu_h = 5.4 \times 10^{-10} n_n T^{\frac{1}{2}}$. Here T is the electron temperature in Kelvin and $n_n$ is the neutral density (number per cubic cm). The neutral density values and the temperature required for the calculation of the collision frequency are obtained from the NASA MSIS90 model and IRI model respectively. We have run the LWPC RTABLE module throughout the day both in the presence and absence of the flares to find the VLF amplitude variation with adjusted temporal separation.   \\

\section{Results}

\begin{figure}
  \centering
  \includegraphics[scale=0.32]{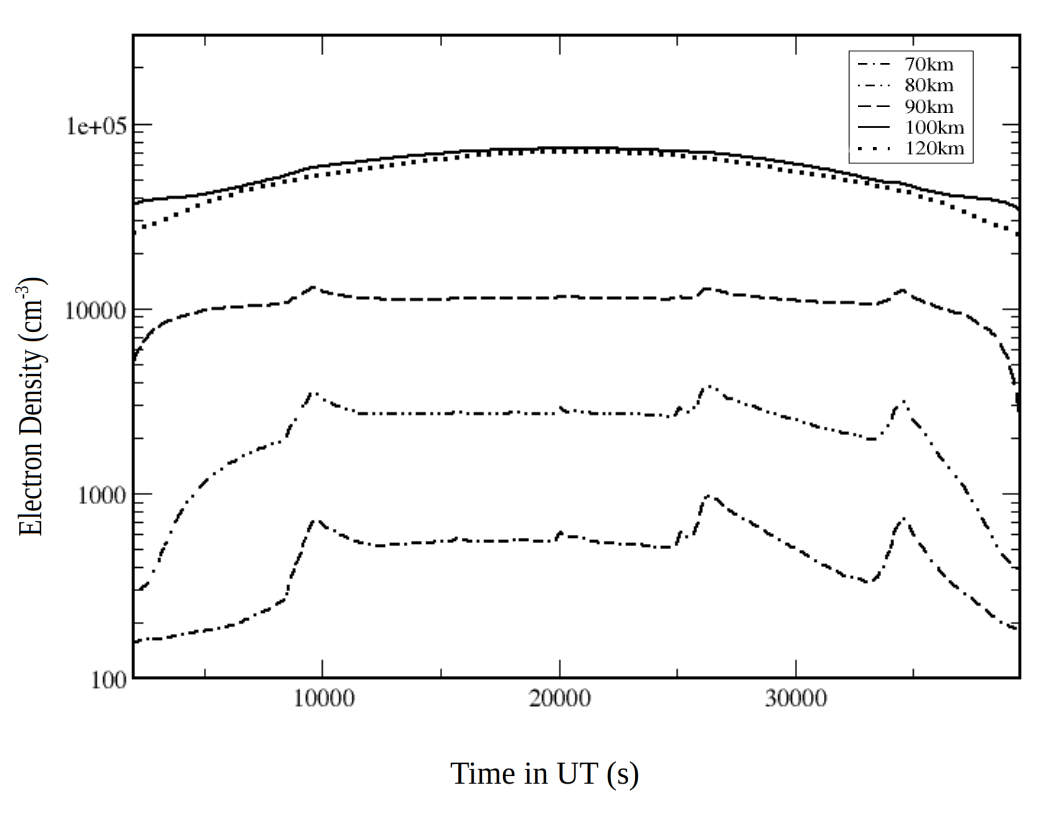}
  \caption{Simulated electron density variation as a function of time throughout the day time at various altitudes over the receiver location for the day of analysis}
  \label{edentot}
\end{figure}

\noindent The Figure~\ref{edentot} demonstrates the electron density variation as function of time throughout the day at different altitudes, calculated 
with the ion-chemistry model using the computed total ionization rates at different heights. We can see that in the range of altitude from $\sim$ 70 -- 100 km the electron density peaks during the flares, with maximum increment around 90 km. Above this height though the rates of ionization increases during flares, the relative increment in electron density with respect to the pre-flare electron density values are small and around and above $\sim$ 120 km  there are almost negligible relative increase in electron density during flares. \\

\begin{figure}[ht]
  \centering
  \includegraphics[scale=0.22]{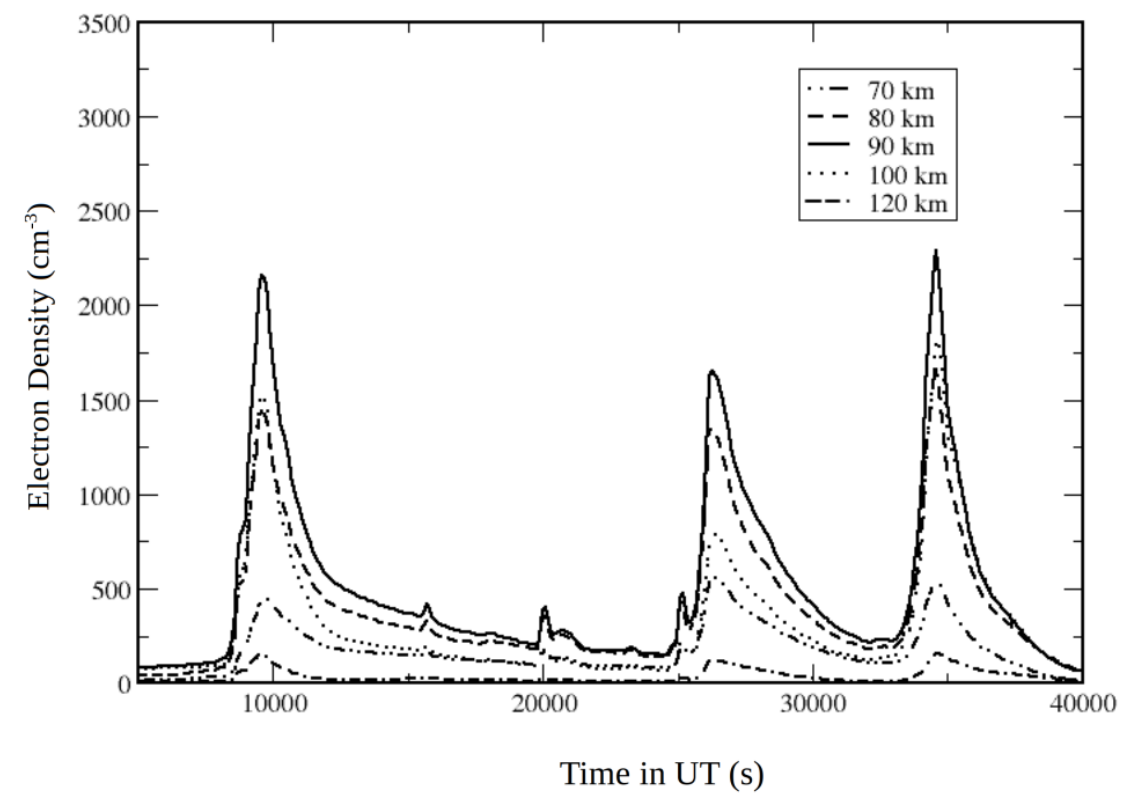}
  \caption{Simulated electron density variation added by the flares above the ambient density as function of time through the whole day time at various altitudes over the receiver location}
  \label{edenflare}
\end{figure}

\noindent This is further illustrated in Figure~\ref{edenflare}, which depicts the excess electron density above ambient levels at all those heights caused by the presence of flares throughout the day. To obtain these we repeat the process of calculating the electron density values assuming there was no flare occurring that day. We systematically remove the X-ray flux and EUV photons due to the flares from the light curves demonstrated in Figure~\ref{xraylc} and Figure~\ref{uvlc} respectively and fitted the base flux or photon levels with smooth lines. Then we use these base light curves in our codes to calculate the ionization rates at all heights for the two sources, sum them up and then compute electron density at different heights throughout the day. In the next step we subtract the electron density values calculated from non-flare assumption from the corresponding electron density values obtained in presence of flare (demonstrated in  Figure~\ref{edentot}) for different altitudes. We plot the resultant electron density values as function of time for various heights in  Figure~\ref{edenflare}. Though the Figure here indicates that at $\sim$ 90 km  the increase in electron density contributed by the flares is the highest, as we have calculated electron density values at 2 km intervals we find through basic interpolation that at around 92.5 km the electron density increases maximum. Below and above this height the increase in electron density gradually becomes lower. \\

\begin{figure}
  \centering
  \includegraphics[scale=0.21]{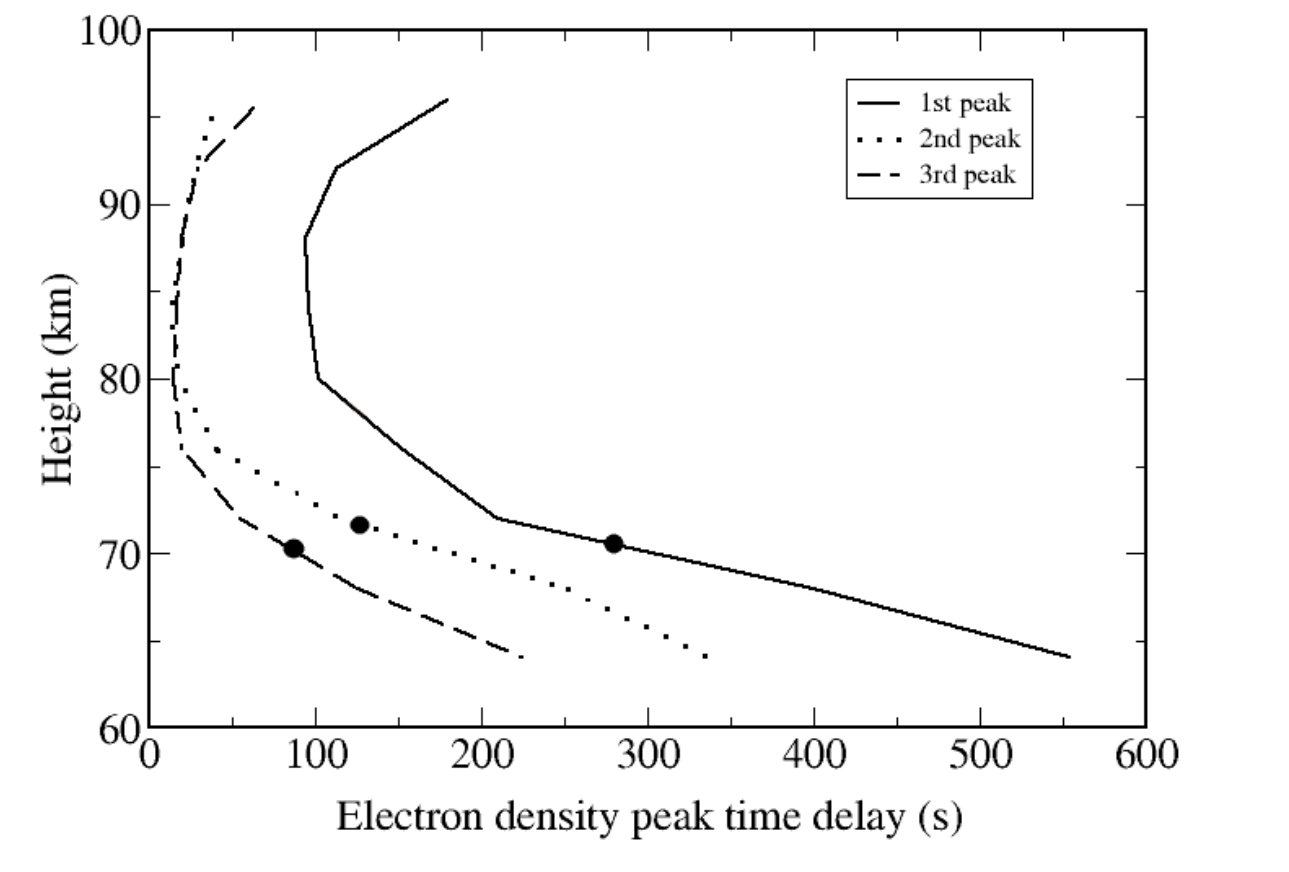}
  \caption{Peak time delay between the soft X-ray peak and computed electron density peaks at different heights for the three M-class flares occurring during the day are shown. Corresponding peak time delay of the observed VLF amplitude for the three flares are shown with big black dots.}
  \label{ptd}
\end{figure}

\noindent  Following the analysis presented in \cite{palit15} here we have estimated the electron density peak time delay (PTD), i.e., the delay between soft X-ray peak and the computed electron density peaks at different heights for all the three M-class flares. In Figure~\ref{ptd} we plotted the PTD values for different heights as function of height for all the three flares. In agreement with results those we presented earlier, this time also the PTDs for all the flares are found to decrease as height increases from $\sim$ 60 km up to $\sim$ 80 km and then increase above it. The three large dots represent the PTD between the VLF amplitude peaks and the corresponding flare soft X-ray peaks for all the three flares. These values are $\sim 280$ sec for the first, $\sim 125$ sec for the second and $\sim 87$ sec for the third M-class flare. We draw the PTDs of VLF for all the flares on the plots of the same for the electron density values of the corresponding flares to show that the PTDs of VLF signal match those of electron density for all the flares between $\sim$ 70 - 72 km altitude. Following the suggestion from \citet{palit15} we can infer that during the peak of the flares the VLF radio waves follows the timing profile of electron density at those particular heights and the effective VLF reflection heights ($h_{eff}$) proposed  by \cite{palit15} reside in the above mentioned altitude range. \\

\begin{figure}[ht]
  \centering
  \includegraphics[scale=0.32]{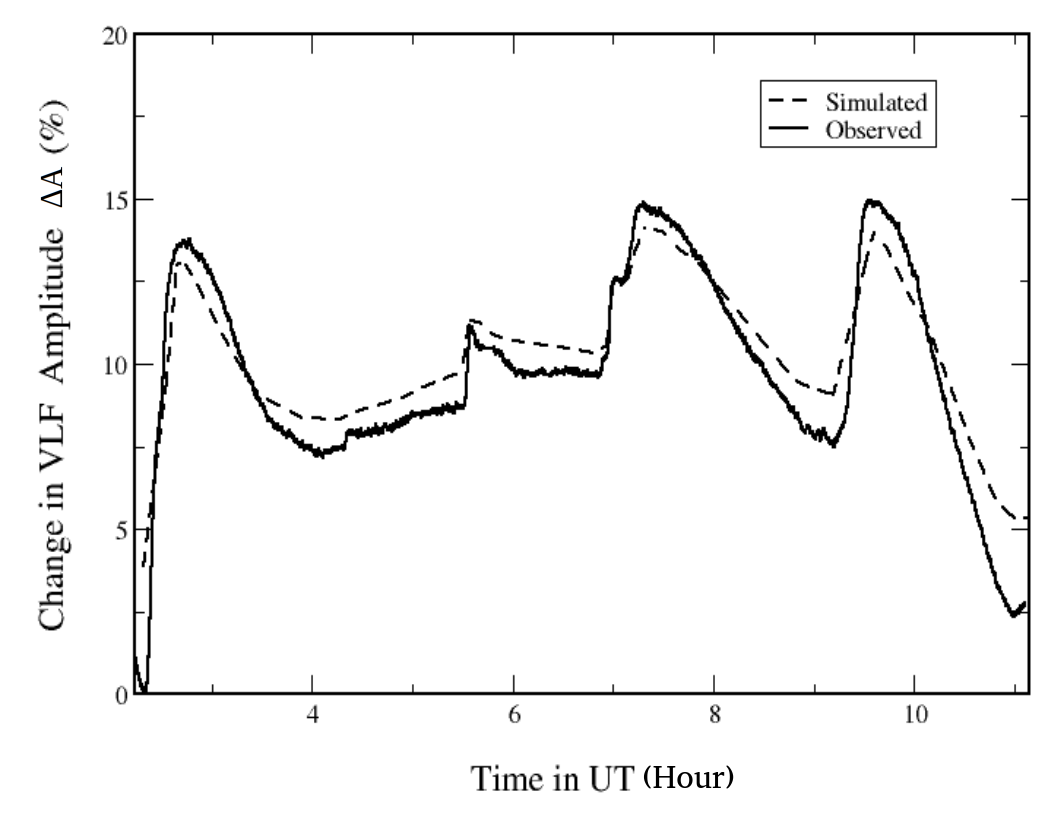}
  \caption{Simulated VLF amplitude modulation (\% change from the amplitude value at the start of the day) is plotted with the observed VLF amplitude \% change.}
  \label{modamp}
\end{figure}

\noindent In Figure~\ref{modamp} we plotted the change in VLF amplitudes (in \%) from the values at the start of the day, when the amplitude starts to rise during morning terminator for both the observed VLF signal and simulated VLF signal. The observed VLF amplitude is plotted for each second. We calculated and plotted the simulated amplitude values at intervals, which are not regular but effectively demonstrate the varying rates of modulation during both flare and non-flare periods. Obviously the simulated plot can not capture the smallest of the variations in the signal amplitude, but it seems to follow the observed signal amplitude variation quite closely. The peak times of the simulated VLF amplitudes are found to closely match, if not exactly coincide with those of the observed VLF signal amplitudes for all the flares.  \\

\section{Discussions and Conclusions}
\noindent In this study, we present a numerical reconstruction of the observed VLF amplitude modulation over an entire day influenced by multiple solar flares. The reconstruction process simulates the underlying physical and chemical interactions, beginning with the calculation of ionization rates in the lower ionosphere due to high-energy solar photons, performing the ion-chemical evolution involving the neutral atoms and molecules and finally extending to the response of the VLF signal to the resulting enhancements in plasma density within that region. The elevated plasma density in the lower ionosphere by the extra ionizing radiation produced by the flares leaves its signature on the VLF sky wave by increasing absorption of wave energy and modulation of effective reflection heights. Here the combined effect is found to be increase in VLF amplitude during the occurrence of the flares. Instead of conventional approach of analyzing such effect by empirically finding the modulation in Wait's parameters and hence probable electron density modulation, we take the forward approach of starting from the incident soft-X-rays and EUV photons and their ionization and finally finding the VLF amplitude modulation. The work can be considered as the extension of similar work done by \citep{palit13}, where the reconstruction of affected VLF amplitudes were done for two stand alone large solar flares. In contrast, the present study addresses the added complexity arising from modulation of the background plasma environment by successive solar flares.\\

\noindent For this particular work we are dealing mainly with multiple large (M-class) flares occurring through a whole day time over the chosen GCP. In earlier work we dealt with single isolated flare each time and the calculation of electron-ion density enhancement with ion-chemistry code required input of initial density values at different lower ionospheric heights from an empirical model, namely the IRI model. Here, we realize that the initial plasma condition found by any flare other than the first one should be modified by the previous flares. So, treating them individually with incorporating initial plasma values from an empirical model each time would not work. So, we do not deal with each of the flares individually, rather run the ion-chemistry model only once at the start of the day, i.e., the sunrise terminator and continue throughout the day using initial concentration values from the IRI model only once at the beginning of the model run.\\

\noindent  In earlier work, only individual flares were used and that too without using EUV in the calculation of ionization with the realization that the flare induced EUV should have negligible effect compared to enhanced soft X-ray photons. In the current scenario the VLF amplitude modulations due to the flares can be considered as superimposed on the diurnal profile of the VLF amplitude profile had there been no flares. This ambient diurnal variation is produced by the continuous solar EUV photon incidence and ionization in the lower ionosphere and is modulated by the solar zenith angle variation over a day. Since our goal is to numerically reproduce the complete diurnal variation of VLF amplitude, including the effects of multiple solar flares, it is essential to incorporate solar EUV-induced ionization into our calculations. As expected we find that the diurnal profile of the VLF amplitude has been satifactorily generated by the simulation as EUV ionization is integrated in it, though the effect of these ionization is almost negligible compared to the soft X-ray ionization during the flares. \\  

\noindent There are several sources of uncertainty and error in such simulation, such as the possible inaccuracy in available chemical reaction cross-sections of the D-region ion-chemistry, inability of accounting for all possible irregularities imposed by various local perturbative phenomena, limitation in accuracy of the radio wave propagation code used in the simulation and so forth. In addition to that we have to deal with the fact that after the decommission of RHESSI Solar satellite there are no existing instrument providing us the required X-ray spectra of the solar flares in regular basis. We have to adopt approximation to account for that deficiency. We acknowledge that the assumption in instantaneous X-ray spectral calculation, and the resulting neglect of spectral slope variation over time, may introduce additional error into the results but still the reconstructed VLF amplitude profile quite satisfactorily compares with the observed profile. \\

\noindent In continuation of the work we plan to perform modeling more complex such scenarios such as overlapping flares of different classes at different phases of the solar cycle etc. and improve our model by more accurate flare spectral approximation and more refinement of ion-chemistry model. Such complex modeling will further enhance our quantitative understanding of the effects of ionizing phenomena on the ionosphere and radio wave propagation. It will also support long-term efforts in monitoring and potentially nowcasting space weather using a VLF network integrated with machine learning techniques.\\

\section*{Acknowledgments}
\noindent All the authors acknowledge the Indian Centre for Space Physics (ICSP), Kolkata for providing VLF data.
We acknowledge SOHO and NOAA satellite agency for providing open data sources on solar UV and X-ray data respectively. Special thanks are extended to
Mr. Debashis Bhowmick (ICSP) for his technical and scientific support related to the VLF data.\\

\bibliographystyle{jasr-model5-names}
\biboptions{authoryear}
\bibliography{refs}


\end{document}